\documentstyle[12pt]{article}
 at 14.4truept at 12.0truept
\title{ Improved heavy quark potential at finite temperature
from anti-de Sitter supergravity}
\author{
{\large Satchidananda  Naik} \thanks{e-mail: naik@mri.ernet.in}
\\
  Mehta Research Institute of
 Mathematics \\
 and Mathematical Physics \\
 Chhatnag Road, Jhusi  \\
Allahabad-211 019, INDIA\\}

\begin{document}
\maketitle

\hspace*{\fill}

\hspace*{\fill}
\newcommand{\bee}{\begin{equation}}
\newcommand{\nn}{\nonumber}
\newcommand{\ee}{\end{equation}}
\newcommand{\ba}{\begin{array}}
\newcommand{\ea}{\end{array}}
\newcommand{\bea}{\begin{eqnarray}}
\newcommand{\eea}{\end{eqnarray}}
\newcommand{\ki}{\chi}
\newcommand{\eps}{\epsilon}
\newcommand{\pa}{\partial}
\newcommand{\lb}{\lbrack}
\newcommand{\Se}{S_{\rm eff}}
\newcommand{\rb}{\rbrack}
\newcommand{\de}{\delta}
\newcommand{\th}{\theta}
\newcommand{\rh}{\rho}
\newcommand{\ka}{\kappa}
\newcommand{\al}{\alpha}
\newcommand{\bt}{\beta}
\newcommand{\si}{\sigma}
\newcommand{\bsi}{\Sigma}
\newcommand{\vp}{\varphi}
\newcommand{\gm}{\gamma}
\newcommand{\gb}{\Gamma}
\newcommand{\om}{\omega}
\newcommand{\et}{\eta}
\newcommand{\gt}{ {g^2 T }\over{4 {\pi}^2}}
\newcommand{\qab}{{{\sum}_{a\neq b}}{{q_a q_b}\over{R_{ab}}}}
\newcommand{\omb}{\Omega}
\newcommand{\pr}{\prime}
\newcommand{\ra}{\rightarrow}
\newcommand{\nb}{\nabla}
\newcommand{\MSb}{{\overline {\rm MS}}}
\newcommand{\lnh}{\ln(h^2/\Lambda^2)}
\newcommand{\cz}{{\cal Z}}
\newcommand{\h}{{1\over2}}
\newcommand{\Lm}{\Lambda}
\newcommand{\inft}{\infty}
\newcommand{\abschnitt}[1]{\par \noindent {\large {\bf {#1}}} \par}
\newcommand{\subabschnitt}[1]{\par \noindent
                                          {\normalsize {\it {#1}}} \par}
%-----------------------------------------------------------------------
% The definition below makes spaces e.g \skipp{3} makes 3 spaces
\newcommand{\skipp}[1]{\mbox{\hspace{#1 ex}}}
 
%
%
% various slashed symbols
%
%
%\newcommand\slash#1{\rlap{$#1$}/} % slashes a character
\newcommand\dsl{\,\raise.15ex\hbox{/}\mkern-13.5mu D}
    % this one can be subscripted
\newcommand\delsl{\raise.15ex\hbox{/}\kern-.57em\partial}
\newcommand\Ksl{\hbox{/\kern-.6000em\rm K}}
\newcommand\Asl{\hbox{/\kern-.6500em \rm A}}
\newcommand\Dsl{\hbox{/\kern-.6000em\rm D}} %roman D
\newcommand\Qsl{\hbox{/\kern-.6000em\rm Q}}
\newcommand\gradsl{\hbox{/\kern-.6500em$\nabla$}}
%--------
 %---------------------------------------------------------------
 %\vskip5.0cm
\newpage
\begin{abstract} \normalsize
 We improve the heavy quark potential, extracted from the Wilson loop
average in the Ads/CFT approach  by taking the quantum fluctuation of
the only radial coordinate of $Ads_5$ which is transverse to
the world-sheet of the classical Nambu-Goto string in the static gauge,
and obtain the universal  L{\"u}scher-Symanzik-Weisz/L{\"u}scher term. 

\end{abstract}

\vskip10.0cm
   
\newpage
\pagestyle{plain}
\setcounter{page}{1}
%\setcounter{section}{1}
%\abschnitt{1. Introduction}
%\noindent{\it Introduction}

A seminal conjecture by Maldacena \cite{Mald} states the duality
between the  large $N$ superconformal Yang-Mills theory and the low
energy behaviour of the superstring theory in the Anti-de Sitter space
back-ground. It is further shown that the correlation functions of the 
local gauge  invariant operators of  ${\cal N} = 4$ supersymmetric
Yang-Mills theory are related to the classical 
limit of the superstring partition function
\cite{Pol,Witt}. Also the vacuum expectation values of the non-local operators
like Wilson loops are easily computed using this duality \cite{Mald2,SR}.
This 
facilitates to extract the heavy quark potential for SUSY Yang-Mills theories.
The proposal became quite useful after Witten \cite{Witt2} has shown how one
can break Supersymmetry with non-zero temperature. This makes a close contact
with the pure QCD  and as far as possible, several qualitative 
behaviour of QCD in the strong coupling limit  are verified \cite{qcd}. 
In this approach the background Euclidean time is compactified on a circle
which is related to the equilibrium temperature T. Here  one takes 
the
space-time fermion to be anti-periodic in this direction and  then all the
fermions will be as heavy as the order of the temperature T. In the
extreme high temperature limit fermions   decouple. Thus the d+1
dimensional theory is described as  a  d dimensional theory without any 
fermion. This gives us a d dimensional pure QCD in the zero temperature.

The heavy quark potential of the planar $QCD_3$  is extracted from the 
vacuum expectation value of the Wilson loop operator \cite{SR,BISY}. The
space-like 
Wilson loop gives a linear potential which correctly signals the area law.
However
the  L{\"u}scher-Symanzik-Weisz (LSW) term or commonly known as
L{\"u}scher 
term is absent from this  potential \cite{LSW}. This is a universal term
independent of any coupling constant and is a characterstic of the string 
generated QCD potential and has been seen  and  correctly estimated by
 Lattice  simulation \cite{Bali}. This might cast doubt on the 
supergravity approach to study planar QCD. However  there are some
 attempts  by Greensite and Olesen  \cite{GO} and by Dorn and Otto
\cite{OD} to get rid of this problem.  So far there is no concrete
explanation  in this regard. Here we claim to get this LSW term
by considering quantum fluctuations of the classical solution  
 which is an improvement over the  Nambu-Goto solution taken by authors
of Ref. \cite{SR,BISY} . We will discuss all the points in the sequel.

Here we consider the case of planar $QCD_3$ which can be trivially
extended to higher dimension. Following Maldacena \cite{Mald2}
we put heavy quark and anti-quark at $x = L/2$ and $x = - L/2$
respectively which are attached by a Nambu-Goto string in 
an $Ads_5 \times S_5$  black-hole background, where the metric reads
\bee
ds^2=\alpha'\left\{\frac{U^2}{R^2}\left((1-U_T^4/U^4)dt^2+
  \sum_i dx_i^2\right)+\frac{R^2}{U^2}\frac{dU^2}
{1-U_T^4/U^4}+R^2d\Omega_5^2\right\},
\ee
Where $R^2~ = ~\sqrt{4 \pi g_s N}~ = ~\sqrt{4 \pi {g^2}_{YM} N}$ , and
$U_T = \pi R^2 T$ . Here $g_s$ is the string coupling and $N$ is the
number
of coinciding $D_3$ branes which gives $U(N)$ Yang-Mills theory and $R$ is
the radius of the anti-de Sitter space. The Nambu-Goto string action in
this background is
\bee
S=\frac{1}{2\pi\al}\int d\tau d\sigma \sqrt {\mbox{det}(G_{MN}
\partial_{\mu}x^M\partial _{\nu}x^N)}.
\ee
We look for the Wilson loop which traces area in the space-like surface.
 The boundaries  are accordingly
$x^1=\pm \frac{L}{2},~~~x^0=x^3=0$ and the line where
 $U\rightarrow \infty$
runs parallel to $x^2$ axis. In the static gauge the world-sheet
cordinates
are $\tau ~=~x^2,~~~~\sigma ~=~x^1~$ and $\partial _{\tau}x^M=\delta _2^M$
and $\partial _{\sigma}x^0 = \partial _{\sigma}x^3 = 0 $.
Thus the action 
\bee
S = \frac{\cal T}{2\pi}~\int _{-\frac{L}{2}}
^{+\frac{L}{2}}d\sigma 
\sqrt{\frac{U^4}{R^4}~+~U^{\prime\,2}\frac{U^4}{U^4-U_T^4}}.
\ee
The potential is extracted in the  ${\cal T}\rightarrow \infty$ limit as
\bee
V (L) = {{Lim}_{{\cal T}\rightarrow \infty}}\frac{1}{\cal T} S_{cl}.
\ee
The  ${\cal T}\rightarrow \infty$ limit 
is the analouge of $\hbar \rightarrow 0$ limit and thus the classical
solution is an exact solution. However L{\"u}scher term is a purely
quantum mechanical term showing the finite size scaling. 
 In this very brief report we  show here how this  classical
potential is extracted and how it  can be further  improved  to
accommodate  the
LSW/L{\"u}scher  term.

Using symmetry and the classical equation of motion we get
\bee
\sqrt{\frac{U^4}{R^4}~+~U^{\prime\,2}\frac{U^4}{U^4-U_T^4}} = C_0
~\frac{U^4}{R^4} ~,
\ee

Where $C_0$ is a constant which is fixed from  the boundary  relation 
$U( {\si}_0) = U_0$ is such that $U^{\prime}( {\si}_0) = 0$.
This gives
\bee
\frac{L}{2}-\sigma=\frac{R^2}{U_0}\int_{U/U_0}^\infty
\frac{dy}{\sqrt{(y^4-1)
(y^4-1+\epsilon)}},
\ee
where $\epsilon~ =~ 1-U_T^4/U_0^4$ and the dimension less variable
$y (\si)~=~\frac{U(\si )}{U_0}$.
 The classical action is
\bee
S = \frac{\cal T}{2\pi}~U_0~  \int_{1}^\infty ~ dy~
\frac{ y^4 }{\sqrt{(y^4-1)(y^4-1+\epsilon)}},
\ee
which gives 
\bee
 V( L) ~  =~ \frac{U_0^2}{2\pi R^2}L+\frac{U_0}{\pi}
\int_1^\infty dy\left(\sqrt{\frac{y^4-1}{y^4-1+\epsilon}}-1\right)
+\frac{U_0-U_T}{\pi}.
\ee
 This is a linear potential $V( L) ~= \sigma L +\cdots $
where $\sigma ~ = ~ \frac{U_0^2}{2\pi R^2}$ is the string tension. In
terms of Yang-Mills coupling this gives $\sigma = \frac{\pi}{\sqrt2}
g_{YM}{\sqrt{N T^3}}$. However the subleading term {\bf  $\frac{c}{L}$ }is
missing in the classical approximation. Also {\bf $c$ }is supposed to be
universal term which does not contain any physical parameter except the
finite size scale $L$. This is a purely  quantum effect. 
We take here one-loop quantum effect around the classical solution of $U$
as $U ~=~U_{cl} + \et$ where $\et $ is the quantum field. The action in
this order is given by
\bee
S_2 =  \frac{\cal T}{2 \pi}\int _{\si_0}^{L/2} {\cal L}_2
\ee 
where 
\bee
{\cal L}_2 = \frac{{\pa}^2 {\cal L}}{\pa U^2}~{\et}^2 ~+~2  \frac{{\pa}^2
{\cal L}}{\pa U \pa U^{\pr}}~ \et {\pa \et} ~   
+~ \frac{{\pa}^2 {\cal L}}{\pa{U^{\pr}}^2}~
{(\pa \et)}^2.
\ee
Here ${\cal L}$ is the Lagrangian given in eq.(3).
To integrate over the field $\et$ we redefine  
\bee
\vp = \al ~ \et,
\ee
where $\al $ is a function of the  classical field 
$U_{\it cl}$ obeying equations of motion. (Henceforth we 
denote $U$ as $U_{\it cl}$ and drop the subscript.) 
Now $\et$ is substituted as $\frac{\vp}{\al}$ in eq.(10)
to give
\bee
{\cal L}_2 =  {(\pa \vp )}^2 ~ - 
~2~ \vp~ {\pa \vp}~ ( \frac{{\al }^{\pr}}{\al} - \frac{ b}{ 2 {\al}^2})
~ +~ ( \frac{{{\al }^{\pr}}^2}{{\al}^2} - b \frac{{\al }^{\pr}}{{\al}^3}
+ \frac{c}{{\al}^2} )~ {\vp}^2 ~ + 
~ {\cal L}_{measure},
\ee
and ${\cal L}_{measure}$ is due to the redefinition of the field $\et$
, where
\bea
{\al}^2 & = & \frac{{\pa}^2 {\cal L}}{\pa{U^{\pr}}^2}  \\
 b  & = & 2 \frac{{\pa}^2 {\cal L}}{\pa U \pa U^{\pr}} \nn \\
 c   & = &  \frac{{\pa}^2{\cal L}}{\pa {U^2} }.
\eea
( Here $\pr$ denotes the derivative with respect to $\si$. )
Any generic term  like $f~\pa \vp ~\vp$  can be 
taken as $ \h ~f~ \vp~ \vp {\mid}_{ bounadry}$  and a term like
 $- \h {\pa f}~\vp~ \vp$
due to integration by parts in the action.
We take here the Dirichlet boundary condition for the fluctuating
field $\vp$ and hence drop the boundary term from the action.
 This gives 
\bee
{\cal L}_2 = ( {(\pa \vp )}^2 ~ + ( \frac{{\al }^{\pr \pr}}{\al} - 
\h \frac{{b}^{\pr}}{{\al}^2}+ \frac{c}{{\al}^2} ) {\vp}^2 ) +
~ {\cal L}_{measure}.
\ee
Taking further derivatives with respect to $\si$ of the equation of motion
\bee
\pa ~ \frac{{\pa}{\cal L}}{ \pa U^{\pr}} =   \frac{ \pa {\cal L}}{\pa U},
\ee
one gets 
\bee
\frac{c}{{\al}^2} - \h \frac{{b}^{\pr}}{{\al}^2} = ~ 2 \frac{U^{\pr
\pr}}{U^{\pr}}~ \frac{{\al }^{\pr}}{\al}~ + ~ \frac{U^{\pr \pr
\pr}}{U^{\pr}} .
\ee
Substituting this 
we get 
\bee
{\cal L}_2 =  {(\pa \vp )}^2 ~ + m^2 (\si)~ {\vp}^2
\ee
where the effective mass like term
\bee
 m^2(\si)~=~  ({\pa }^2 \al ~ + ~ 2 \frac{U^{\pr \pr}}{U^{\pr}}~{\pa \al}~
+ ~ \frac{U^{\pr \pr \pr}}{U^{\pr}}~ \al )\frac{1}{\al}~  .
\ee
From eq.(13) $\al$ is given by 
\bee
\al ~=~ \frac{ g (\si )}{U^{\pr }( \si )}
\ee
where 
\bee
g( \si ) ~ = ~ \frac{U_0}{R}~\frac{\sqrt{y^4-1}}{y^2}.
\ee 
This shows that 
\bee
m^2 (\si)~~=~~ \frac{{\pa}^2 g}{g}.
\ee
From eq.(6) one observes that $y(\si)$ remains infinitesimally closer to
1 and then rapidly grows to infinity. If we take $y(\si)~=~1~+~\de (\si)$
where $\de$ is a very small parameter and keeping up to first order
we get
\bee
\frac{L}{2}-\sigma ~= ~const.\sqrt{\de (\si)}
\ee
and for $y (\si )$ little bigger we can neglect 1 compare to $y^4$ in
eq.(6) and get
\bee
\frac{L}{2}~ -~ \sigma ~= ~const.\frac{1}{y^3} .
\ee
For the first case (c.f. eq.(21) and eq.(23))
\bee
g(\si ) ~ =~ p~ ( 1 - \si)
\ee
where $p~=~ 2~ U_0^3~ \frac{\sqrt{\epsilon}}{R^3}$
which gives ${\pa}^2 g = 0 $ and for $y$ large $g(\si)$
is a constant and ${\pa}^2 g = 0 $. This shows that
$m^2(\si )$ is zero for the entire range of integration.

 The measure is
now changed from $D\et ~ \rightarrow ~ D\vp$ and 
 the contribution of this to the  effective action 
\bee
S_{measure}~~ = ~\frac{1}{L}~ \int _{{\si}_0}^{L/2}~ d\si~ ln~ {\al}^2 ,
\ee 
 This integration can be trivially done which contributes a constant 
to the potential.
We take the quantum fluctuation $\et$ to be zero on the boundary of the
Wilson loop i.e.  for $\si = 0$ and $L/2$. 
Then the  integration over $\vp$ gives the most  required LSW term
 {\bf  $ - ~\frac{\pi}{12 L}$  }for the potential which is our main
result.

As mentioned earlier the classical action is exact in the static gauge
when ${\cal T} \rightarrow \inft$. We just supplemented this classical
action with the quantum fluctuation of the transverse cordinate $U$. 
We do not take the fluctuations  of the world-sheet for all the transverse
coordinate  as has been shown by 
 Greensite and Olesen \cite{GO2}. In the finite temperature case and
the high temperature limit when dimensional reduction is taking place,
fermions completely decouple from the spectrum. Since we are in the very
low energy limit of extracting the potential we feel it is legitimate
 not to take any of these fluctuations. 
  % \setcounter{section}{2.0}        
  %  \abschnitt{2. Basic formalism}
\bigskip
   \abschnitt{Note Added }
     This work was completed in December and was presented in the 
    String theory workshop at   Puri (9th Dec- 19th Dec 1998). That time
  I was not aware of the work of Greensite and Olesen \cite{GO2} and also
 of F{\" o}rste et. al.  \cite{FG} where both the groups have  considered
the worldsheet fluctuations of both the fermionic and the bosonic
coordinates. \\
 {\it Acknowledgements:}
 I would like to thank the organisers and all the participants of the Puri
workshop for fruitful discussions.


\newpage
   \begin{thebibliography}{99}

   \bibitem{Mald}
   J. Maldacena, Adv. Theor. Math. Phys. 2 (1998) 231,
     %
    \bibitem{Pol}
     S.S. Gubser, I. Klebanov, A.M.Polyakov 
     Phys. Lett. {\bf B 428} (1998)105    
     % 
   \bibitem{Witt}
   E. Witten, Adv. Theor. Math. Phys.{\bf 2} (1998) 253,
   %
   \bibitem{Mald2}
    J. Maldacena, Phys. Rev. Lett.{\bf 80 }(1998) 4859,
    %
   \bibitem{SR}
   S-J. Rey and J-T. Yee, hep-th/9803001; \\
   S-J. Rey, S. Theisen, and J-T. Yee, Nucl. Phys.{\bf B 527 }(1998) 171
   %
   \bibitem{Witt2}
    E. Witten, Adv. Theor. Math. Phys.{\bf 2} (1998) 505,
    %
    \bibitem{BISY}
   A. Brandhuber, N. Itzhaki, J. Sonnenschein, and S. Yankielowicz,\\ 
   J. High Energy Phys. 06 (1998) 001;  Phys. Lett.{\bf B 434} (1998) 36,
   %
   \bibitem{qcd}
     D. J. Gross and H. Ooguri, Phys. Rev.{\bf D 58} (1998) 106002,\\  
      C. Csaki, H. Ooguri, Y. Oz, and J. Terning, hep-th/9806021\\
     R. de Mello Koch et al. Phys. Rev {\bf D 58} (1998) 105009.
    %
   \bibitem{LSW}
    M. L\"{u}scher, K. Symanzik, and P. Weisz, 
    Nucl. Phys.{\bf B173 }(1980) 365\\
     M. L\"{u}scher, Nucl. Phys.{\bf B190} [FS2] (1981) 31.
    %
    \bibitem{Bali}
      G. Bali, C. Schlichter, and K. Schilling,
     Phys. Rev.{\bf D51} (1995) 5165.

      %
      %   
      \bibitem{GO}
    J. Greensite and P. Olesen, J. High Energy Phys {\bf 08 }(1998) 009,
%
     \bibitem{OD}
     H. Dorn, H.J. Otto , J. High Energy Phys 09 (1998) 021.
      %%
     \bibitem{GO2}
       J. Greensite and P. Olesen,{\it Worldsheet fluctuations and heavy
      quark potential in the Ads/CFT approach} hep-th-9901057   
     \bibitem{FG}
     R.Kallosh and A.A.Tseytlin , J. High Energy Phys 10 (1998) 016\\
     S. F{\"o}rste, D. Ghoshal, S.Theisen,  {\it Stringy corrections to
    Wilson loop in N =4 Super Yang-Mills theory} hep-th- 9903042 
%%%%%%%%%%%%%%%%%%%%%%%%%%%%%%%%%%%%%%%%%%%%%%%%%%%%%%%%%%%%%%%%%%%%%%%
 \end{thebibliography}
 \end{document}